\documentclass[11pt]{article}
\usepackage{charter} 
\usepackage{fullpage}
\usepackage{titlesec}
\usepackage{psfrag}
\usepackage{float}
\usepackage{bbm}
\usepackage{algorithm}
\usepackage[hidelinks]{hyperref}       
\usepackage{algpseudocode}
\usepackage{url}            
\usepackage{booktabs}       
\usepackage{amsfonts}       
\usepackage{nicefrac}       
\usepackage{bm}         
\usepackage{graphicx}		
\usepackage{amsmath} 
\usepackage{diagbox}
\usepackage{enumitem}
\usepackage{siunitx}
\usepackage{multirow}
\usepackage{tabularx}
\usepackage{hhline}
\usepackage{arydshln}		
\usepackage{xcolor}
\usepackage{mathtools} 
\usepackage{amsthm}			

\definecolor{lightgreen}{rgb}{.9,1,.9}

\newcolumntype{L}[1]{>{\raggedright\arraybackslash}p{#1}}
\newcolumntype{C}[1]{>{\centering\arraybackslash}p{#1}}
\newcolumntype{R}[1]{>{\raggedleft\arraybackslash}p{#1}}

\theoremstyle{plain} 

\def\defn{\,\coloneqq\,}
\def\Fix{{\mathsf{Fix}}}

\def\prox{{\mathsf{prox}}}

\newcommand{\R}{\mathbb{R}}

\def\Dbm{{\bm{D}}}

\def\defn{\,\coloneqq\,}
\def\Fix{{\mathrm{Fix}}}

\def\prox{{\mathrm{prox}}}

\def\R{\mathbb{R}}

\def\ebm{{\bm{e}}}

\def\xbm{{\bm{x}}}
\def\zbm{{\bm{z}}}
\def\ybm{{\bm{y}}}

\def\zbm{{\bm{z}}}

\def\wbm{{\bm{w}}}

\def\Abm{{\bm{A}}}

\def\Hbm{{\bm{H}}}
\def\Sbm{{\bm{S}}}
\def\Dbm{{\bm{D}}}
\def\Pbm{{\bm{P}}}
\def\Fbm{{\bm{F}}}
\def\Ibm{{\bm{I}}}

\def\thetabm{{\bm{\theta }}}
\def\Abm{{\bm{A}}}
\def\Dbm{{\bm{D}}}
\def\Pbm{{\bm{P}}}
\def\Sbm{{\bm{S}}}
\def\Fbm{{\bm{F}}}
\def\Ibm{{\bm{I}}}

\def\Dsf{{\mathsf{D}}}

\def\Isf{{\mathsf{I}}}

\def\Tsf{{\mathsf{T}}}
\def\Tsf{{\mathsf{T}}}

\def\Dsf{{\mathsf{D}}}

\def\Isf{{\mathsf{I}}}

\def\xbmast{{\bm{x}^\ast}}
\def\xbmbar{{\overline{\bm{x}}}}

\def\argmin{\mathop{\mathrm{arg\,min}}} 

\usepackage{upgreek}
\usepackage[numbers,sort&compress]{natbib}
\title{Robustness of Deep Equilibrium Architectures to\\Changes in the Measurement Model}

\author{
Junhao Hu$^{\dagger}$ , Shirin Shoushtari$^{\dagger}$, Zihao Zou,\\ Jiaming Liu, Zhixin Sun, and Ulugbek S.~Kamilov\thanks{This material is based upon work supported by the NSF CAREER award under grant CCF-2043134.}\\
\emph{\footnotesize Computational Imaging Group (CIG), Washington University in St.\ Louis, MO, USA}\\
\emph{\footnotesize $^{\dagger}$These authors contributed equally.}
}
\date{}%
\begin{document}
%
\maketitle
\begin{abstract}
\noindent
Deep model-based architectures (DMBAs) are widely used in imaging inverse problems to integrate physical measurement models and learned image priors. Plug-and-play priors (PnP) and deep equilibrium models (DEQ) are two DMBA frameworks that have received significant attention. The key difference between the two is that the image prior in DEQ is trained by using a specific measurement model, while that in PnP is trained as a general image denoiser. This difference is behind a common assumption that PnP is more robust to changes in the measurement models compared to DEQ. This paper investigates the robustness of DEQ priors to changes in the measurement models. Our results on two imaging inverse problems suggest that DEQ priors trained under mismatched measurement models outperform image denoisers.
\end{abstract}
%
%
\section{Introduction}
\label{sec:intro}

\noindent
Many imaging problems---such as image denoising, deblurring, super-resolution, and reconstruction---can be formulated as imaging inverse problems. Deep learning (DL) has become a popular data-driven strategy for solving imaging inverse problems by training deep neural net architectures to map noisy measurements to the desired images~\cite{McCann.etal2017}. Among various DL architectures for inverse problems, deep model-based architectures (DMBAs) have received significant attention due to their ability to integrate physical measurement models and image priors specified as convolutional neural nets (CNN).  Well-known strategies for designing DMBAs include plug-and-play priors (PnP), regularization by denoising (RED), deep unfolding (DU), and deep equilibrium architectures (DEQ) (see review papers~\cite{Lucas.etal2018, Ongie.etal2020, kamilov2022plug}). DMBAs can be systematically obtained from model-based iterative algorithms by parametrizing the regularization step as a CNN and training it to adapt to the empirical distribution of desired images.

\medskip\noindent
 Current DMBA strategies can be conceptually divided into two categories. The first category consists of models that rely on image priors trained independently of the measurement model. PnP and RED are two well-known frameworks that specify image priors using image denoisers trained to remove additive white Gaussian noise (AWGN) in the first category~\cite{venkatakrishnan2013plug, Romano.etal2017, Kamilov.etal2017}. DU and DEQ are two well-known frameworks in the second category where the image prior is trained to be end-to-end optimal for a specific inverse problem~\cite{Aggarwal.etal2019, Monga.etal2021, gilton2021deep}. Since all DMBA categories use the knowledge of the measurement model during inference, it is commonly accepted that DMBAs are more robust than generic CNNs to changes in the measurement model~\cite{Ongie.etal2020}. On the other hand, since the image prior in PnP/RED is independent of the specific measurement model, it is a common assumption that PnP/RED are more robust than DU/DEQ to changes in the measurement models. Despite the rich literature on DMBAs, the robustness of DEQ to changes in the measurement models has never been systematically compared to PnP. 
 
 \medskip\noindent
 This paper addresses this gap by comparing DEQ to PnP. Both frameworks can be viewed as \emph{implicit neural networks} with potentially an infinite number of layers~\cite{gilton2021deep}. While the image priors in PnP are AWGN denoisers, those in DEQ are \emph{artifact removal (AR)} operators trained end-to-end using specific measurement models. We consider two distinct inverse problems: (a) compressive sensing magnetic resonance imaging (CS-MRI)~\cite{lustig2008compressed} and (b) image super-resolution with known blur kernels~\cite{Almeida.Figueiredo2013}. We use the same deep architecture for both PnP and DEQ derived from the well-known model-based iterative algorithms. Our results suggest that contrary to common intuition, AR priors trained using mismatched measurement models within DEQ can perform better relative to the pure AWGN priors in PnP. We observe that mismatched AR priors outperform AWGN priors on average by 1.84 dB in CS-MRI and 0.23 dB in image super-resolution.

\section{Inverse Problems}
\label{sec:backg}
\noindent
Recovering an unknown image $\xbmast \in \R^n$ from its sub-sampled and noisy measurements 
\begin{equation}
\ybm = \Abm \xbmast + \ebm,
\end{equation}
is often formulated as an inverse problem, where $\Abm \in \R^{m \times n}$ is the measurement model that characterizes the response of a physical system and $\ebm$ is the AWGN. The inverse problem is commonly formulated as optimization
\begin{equation}
\label{con:inverse}
\widehat{\xbm} = \argmin_{\xbm \in \mathbb{R}^n} f(\xbm) \quad \text{with} \quad f(\xbm) = g(\xbm) + h(\xbm),
\end{equation}
where $g$ is the data-fidelity term that measures the consistency of the solution with $\ybm$, and $h$ is a regularizer that enforces prior knowledge on $\xbm$. For example, two traditional data-fidelity and regularization terms are the least-squares function $g(\xbm) = \frac{1}{2} \|\ybm - \Abm \xbm\|_2^2$ and the total variation (TV) function $h(\xbm) = \tau \|\Dbm \xbm\|_1$, where $\tau >0 $ is the regularization parameter and $\Dbm$ is an image gradient~\cite{Beck.Teboulle2009a}. 

\medskip\noindent
When the function $g$ or $h$ in~\eqref{con:inverse} is nonsmooth, the optimization problem is often solved using a proximal algorithm. Two widely-used proximal algorithms are the accelerated proximal gradient method (APGM)~\cite{beck2009fast} and the alternating direction method of multipliers (ADMM)~\cite{eckstein1992douglas}. Given a proper, closed, and convex function $g$, both of these algorithms avoid differentiating it by relying on the proximal operator
\begin{equation}
\label{eq:proximaloperator}
\prox_{\gamma g} (\zbm) := \argmin_{\xbm \in \R^n} \bigg\{ \frac{1}{2} \|\xbm -\zbm\|^2_2 + \gamma g(\xbm) \bigg\},
\end{equation}
where the parameter $\gamma >0$ is analogous to step-size. 

\section{Deep Model-based Architectures}

\begin{figure}[t]
\centering
\includegraphics[width=0.8\textwidth]{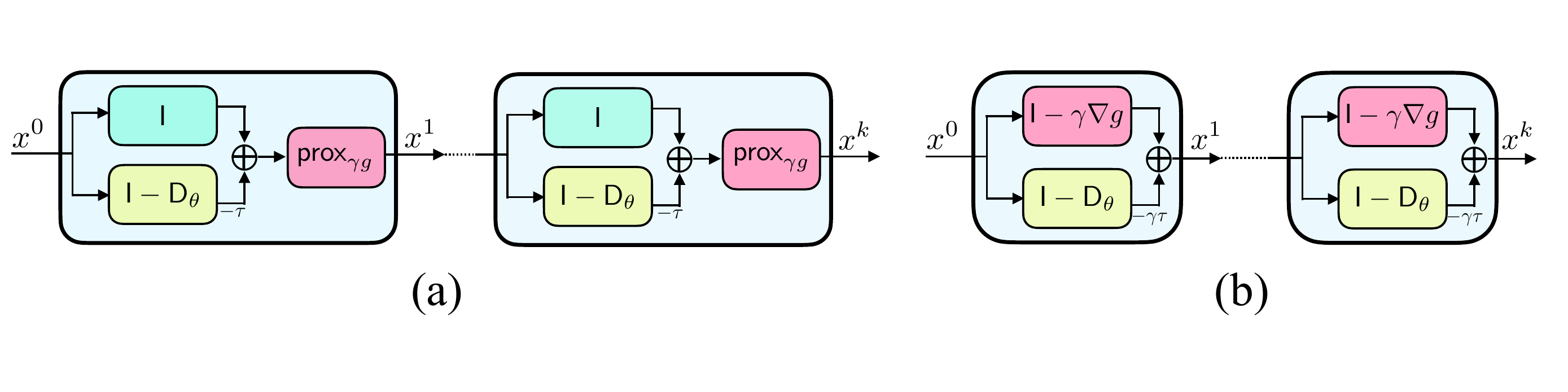}
\caption{\small Schematic illustration of two DMBAs for inverse problems: (a) PnP-PGM and (b) SD-RED. While both architectures have the same set of fixed points, their difference is in the use of the data-fidelity term, namely PnP-PGM uses $\prox_{\gamma g}$ and SD-RED uses $\nabla g$.}
\label{Fig:structure}
\vspace{-.5em}   
\end{figure}

\noindent
We now present the details of DMBAs used in this paper. We first present two deep architectures and then discuss two training approaches for the image priors.

\subsection{Plug-and-Play Architectures}

\noindent
We adopt \emph{steepest descent} variant of RED (SD-RED) and \emph{proximal gradient method} version of PnP (PnP-PGM) for CS-MRI and image super-resolution experiments, respectively. The SD-RED updates can be expressed as
\begin{equation}
\xbm^k =  \Tsf_{\text{SD-RED}}(\xbm^{k-1})\quad \text{with} \quad  \Tsf_{\text{SD-RED}} (\xbm)= \xbm - \gamma(\nabla g(\xbm)+\tau (\xbm-\Dsf(\xbm))),
\label{con:RED}
\end{equation}
 where $\gamma >0 $ is the step-size, $\Dsf$ is the CNN prior, and  $\tau >0$ is the regularization parameter. Note that if the conditions on $\Dsf$ from~\cite{Romano.etal2017} are satisfied then $\tau(\xbm - \Dsf(\xbm))$ is interpretable as a gradient of some convex regularizer $h$. Note that depending on the training procedure (see Section~\ref{Sec:TrainingPriors}), the CNN prior can correspond to an AWGN denoiser or AR operator.

Instead of using the gradient $\nabla g$, PnP-PGM performs a proximal update $\prox_{\gamma g}$ with respect to the data-fidelity term $g$. This update has a closed-form solution for image super-resolution, which makes it preferable for that problem. The update rule for PnP-PGM can be summarized as 
\begin{equation}
    \label{PnPrule}
    \xbm^k = \Tsf_{\text{PnP-PGM}}(\xbm^{k-1})\quad \text{with}\quad\Tsf_{\text{PnP-PGM}}(\xbm)=\prox_{\gamma g}(\xbm - \gamma\tau (\xbm - \Dsf(\xbm))),
\end{equation}
where $\tau > 0$ is the regularization parameter, and $\gamma > 0 $ is the step-size. For linear inverse problems, the proximal operator has the following closed-form solution
\begin{equation}
\label{Prox_GS-PnP}
\prox_{\gamma g} (\zbm) = (\gamma \Abm^\Tsf\Abm+\Ibm)^{-1} (\zbm + \gamma \Abm^\Tsf \ybm),
\end{equation}
which can be efficiently evaluated in the Fourier domain for image super-resolution~\cite{Afonso.etal2010}.

\medskip\noindent
It is straightforward to verify that both methods have the same set of fixed points $\xbmbar \in \Fix(\Tsf) \defn \{\xbm\in\R^n:\Tsf(\xbm)=\xbm\}$ that balance the measurement model and learned prior model.

\subsection{Training Image Priors}
\label{Sec:TrainingPriors}

\medskip\noindent
The traditional PnP strategy considers the following image denoising problem
\begin{equation}
\label{Eq:MAPDenoise}
\zbm = \xbm_0 + \wbm \quad \xbm_0 \sim p_{\xbm_0}\,, \quad \wbm \sim \mathcal{N}(0, \sigma^2 \Ibm)\,,
\end{equation}
and trains $\Dsf$ as a CNN that maps $\zbm$ to $\xbm_0$. Since the training does not use the measurement model $\Abm$, the prior is viewed as a generic image prior usable in multiple applications. 

\begin{figure*}[t]
\centering
\includegraphics[width=0.9\textwidth]{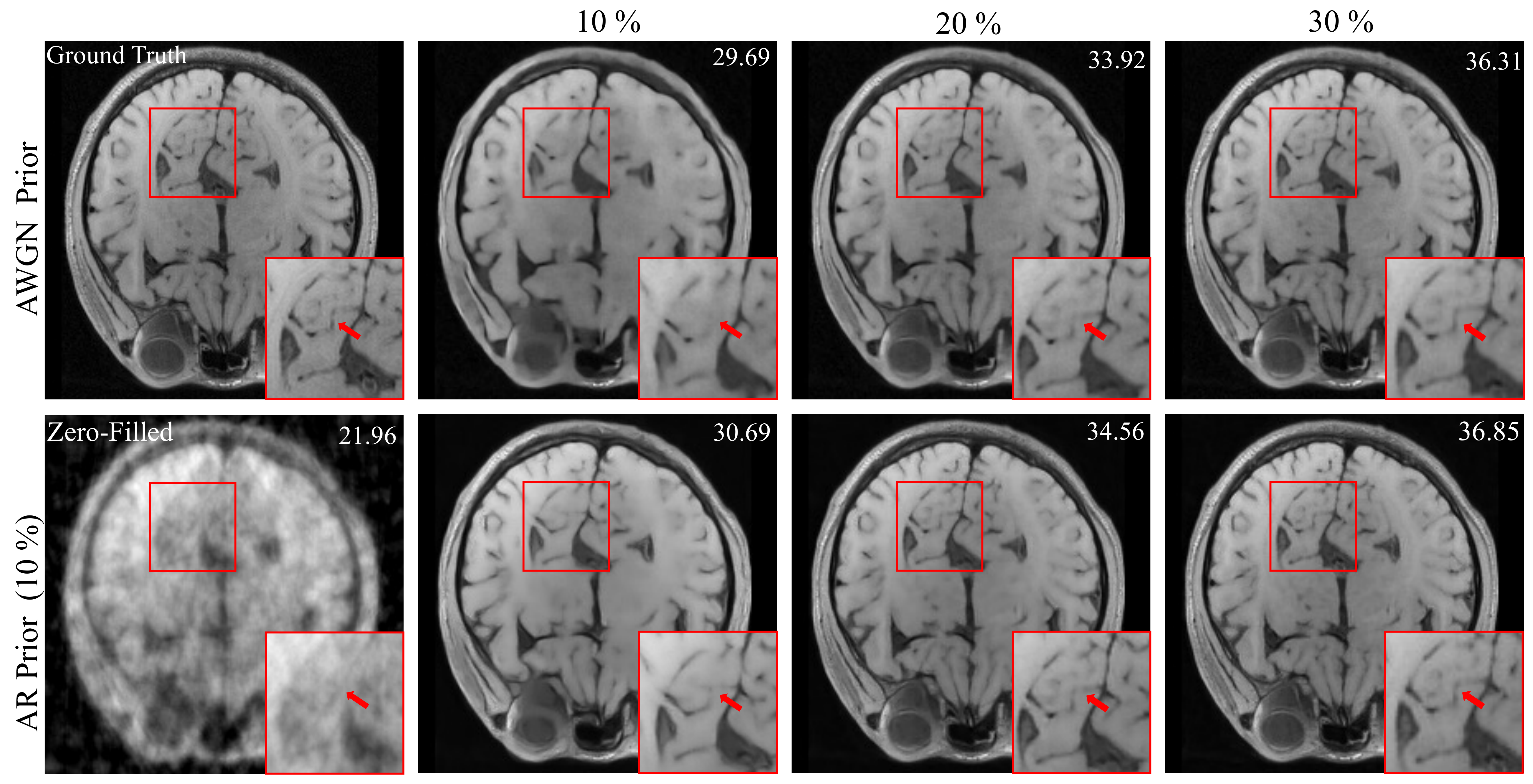}
\caption{\small The comparison of AWGN (PnP) and AR (DEQ) priors on CS-MRI image reconstruction at $10\%$, $20\%$, and $30\%$ radial Fourier sampling. Since the AR prior was trained at $10\%$ sampling rate, it is a mismatched prior for $20\%$ and $30\%$ sampling scenarios. Note how the mismatched AR prior outperforms AWGN prior in every scenario, showing its robustness to measurement model mismatch. 
}
\label{Fig:mri_mismatched_training}
\end{figure*}

\medskip\noindent
DEQ seeks to train $\Dsf$ by minimizing the discrepancy between the fixed-point $\xbmbar = \Tsf_{\thetabm}(\xbmbar)$ and the ground truth image $\xbmast$
\begin{equation}
\label{Eq:DEQ1}
\ell(\thetabm)= \frac{1}{2}\|\xbmbar(\thetabm)-\xbmast\|_2^2.
\end{equation}
The gradient of the loss with respect to $\thetabm$ can be computed using implicit differentiation at the fixed-points 
\begin{equation}
\label{Eq:DEQ2}
\begin{aligned}
\nabla\ell(\thetabm)= (\nabla_\thetabm \Tsf_\thetabm(\xbmbar))^\Tsf \left(\Isf - \nabla_\xbm \Tsf_{\thetabm}(\xbmbar)\right)^{-\Tsf}(\xbmbar-\xbmast),
\end{aligned}
\end{equation}
where $\ell$ is given in~\eqref{Eq:DEQ1} and $\Isf$ is the identity mapping. Since DEQ includes the information of the measurement model $\Abm$ (embedded in the operator $\Tsf$), the corresponding AR operator can be viewed as a problem-specific image prior.

\medskip\noindent
We consider image priors trained via DEQ using a ``mismatched'' measurement operator $\Abm^\prime$ and applied at inference time using the true measurement operator $\Abm$. Accordingly, we train SD-RED by replacing $\nabla g$ in \eqref{con:RED} with $\nabla g^\prime(\xbm) = \Abm^{\prime ^ {\Tsf}}(\Abm^\prime\xbm - \ybm)$. Similarly, PnP-PGM is trained by replacing $\prox_{\gamma g}$ in~\eqref{Prox_GS-PnP} by a mismatched update rule
\begin{equation}
\label{Prox_mismatched}
\prox_{\gamma g'} (\zbm) = (\gamma \Abm^{\prime^\Tsf}\Abm^\prime+\Ibm)^{-1} (\xbm + \gamma \Abm^{\prime^\Tsf} \ybm).   
\end{equation}

\begin{figure*}[t]
\centering
\includegraphics[width=1\textwidth]{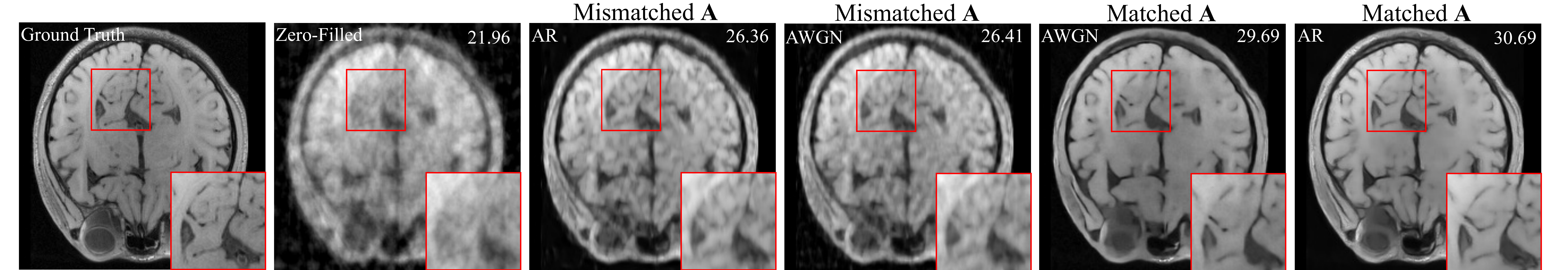}
\caption{\small Illustration of using mismatched measurement models \emph{during inference}. The mismatched setting is obtained by using the measurement model at $20\%$ sampling for reconstructing from data corresponding to $10\%$ sampling. Note the dramatic performance drop due to the usage of mismatched measurement models during inference for both AWGN and AR priors.}
\label{Fig:mri_mismatched_inference}
\vspace{-.5em}   
\end{figure*}


\section{Numerical Results}
\label{sec:NumRes}

\noindent
Our numerical results evaluate the robustness of DEQ priors to changes in the measurement operator $\Abm$ in the context of two imaging problems: CS-MRI and image super-resolution. 

\begin{figure*}[t]
\centering
\includegraphics[width=0.9\textwidth]{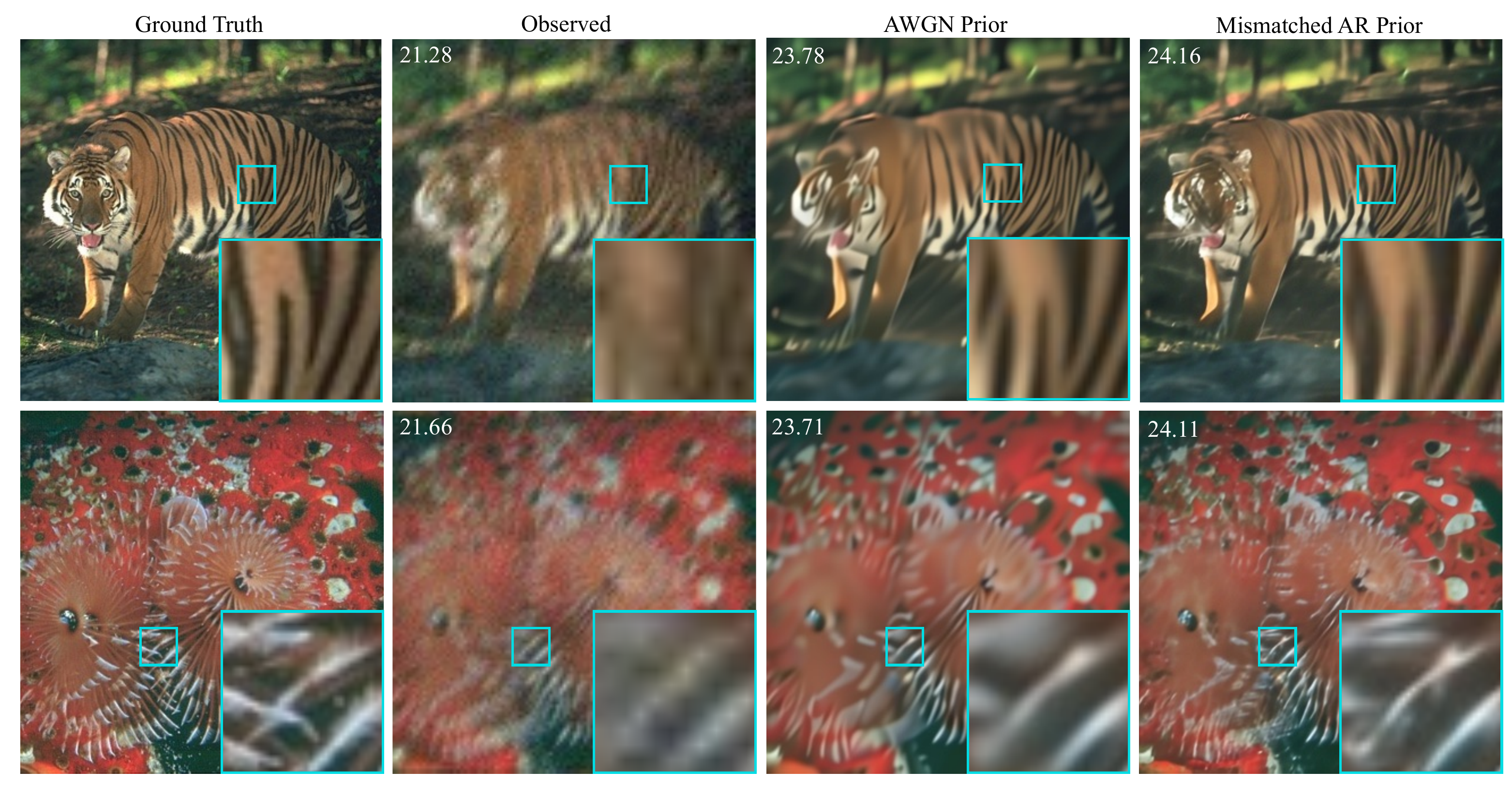}
\caption{\small Image super-resolution on two images using an AWGN prior and a mismatched AR prior. The measurement model corresponds to scaling of $3\times$ superresolution with a input noise level of $\nu=0.03$. Note how the mismatched AR prior outperforms the AWGN prior.}
\label{Fig:blur}
\vspace{-.5em}   
\end{figure*}

\subsection{Compressive Sensing MRI}

\medskip\noindent
In CS-MRI, the goal is to recover an image $\xbmast$ from its sparsely-sampled Fourier measurements. We simulate a single-coil CS-MRI with radial Fourier Sampling. The measurement model $\Abm$ in CS-MRI is $\Abm = \Pbm \Fbm$, where $\Fbm$ is the Fourier transform and $\Pbm$ is a diagonal sampling matrix. 

\medskip\noindent
Image priors were trained using the brain dataset~\cite{zhang2018ista}, where we used 10 slices of 256 $\times$ 256 images as the test images. The AWGN denoisers for PnP correspond to a DnCNN architecture~\cite{zhang2017beyond} trained using noise levels $\sigma \in \{1, 2, 3, 5, 7, 8, 10, 12, 15\}$. For each experiment, we select the denoiser that achieves the highest PSNR. We trained AR operators for DEQ using different CS ratios ($m/n$) with the same DnCNN architecture. The AR priors were initialized using the pre-trained AWGN denoiser with $\sigma = 5$.
Nesterov~\cite{nesterov2003introductory} and Anderson accelerations~\cite{anderson1965iterative} were used in the forward pass and the backward pass during DEQ training.

\begin{table}[t]
\caption{Average PSNR (dB) for CS-MRI.}
    \centering
    \renewcommand\arraystretch{1.2}
    {\footnotesize
    \scalebox{0.96}{
    \begin{tabular*}{301pt}{|L{100pt} | C{35pt} C{35pt} C{35pt}  C{35pt} | }
        \hline
        
        {\textbf{CS ratio}} & {10\%} & {20\%} & {30\% } & {Avg}
         \\ \hline
         {\textbf{AWGN Prior}} &{31.37}&{35.52}&{37.88} & {34.92}\\
         {\textbf{Mismatched AR Prior}} &{\textbf{33.05}}&{\textbf{37.48}}&{\textbf{39.77}} & {\textbf{36.76}}\\
         \hline
        
    \end{tabular*}}
    }
\label{Tab:CS-MRI}
\end{table}

\medskip\noindent
Fig.~\ref{Fig:mri_mismatched_training} shows results on a test image from~\cite{zhang2018ista} at three sampling ratios: $10\%$, $20\%$, and $30\%$. The AR prior in the reconstruction was trained using the measurement model corresponding to  $10\%$ sampling. Therefore, it is a mismatched prior for performing inference at $20\%$ and $30\%$ sampling. Despite the mismatch, the AR prior significantly outperforms the AWGN prior in all considered scenarios. Table \ref{Tab:CS-MRI} reports all the comparisons between the mismatched AR priors and AWGN priors in CS-MRI. Note that the mismatched AR operator outperforms the AWGN prior in all experiments.

\medskip\noindent
 Fig.~\ref{Fig:mri_mismatched_inference} illustrates the impact of using an inaccurate measurement model \emph{during inference}. The results are obtained by using the measurement model at $20\%$ sampling for reconstructing from data corresponding to $10\%$ sampling. Fig.~\ref{Fig:mri_mismatched_inference} shows the results obtained using matched and mismatched measurement models at inference for both AWGN and AR priors. We observe a severe performance drop due to the usage of an inaccurate measurement model during inference, which highlights the importance of the measurement models in DMBAs.

\subsection{Image Super-Resolution}

\medskip\noindent
The measurement model in image super-resolution corresponds to $\Abm = \Sbm \Hbm$, where $\Hbm$ is the convolution with an anti-aliasing kernel, $\Sbm$ is the standard $s$-fold downsampling matrix of size $m \times n$, and $n = s^2 \times m$. The priors were trained using color image dataset in~\cite{zhang2021plug}. CBSD68 dataset proposed in~\cite{martin2001database} was used for inference. Three Gaussian blur kernels (kernels (b), (d), and (e) from ~\cite{hurault2021gradient}) were used to downsample images at scale $s=3$ for inference, and jointly, at scale $s=2$ and $s=4$ during training of the DEQ prior. The CNN priors architectures correspond to U-net~\cite{liu2022online}. 
\begin{table}[t]
\caption{Average PSNR (dB) for image super-resolution.}
    \centering
    \renewcommand\arraystretch{1.2}
    {\footnotesize
    \scalebox{0.96}{
    \begin{tabular*}{301pt}{|L{100pt} | C{35pt} C{35pt} C{35pt}  C{35pt}|}
        \hline
        {\textbf{Blur kernel}} & {kernel 1} & {kernel 2} & {kernel 3} & {Avg}
         \\ \hline
         {\textbf{AWGN Prior}} &{24.63}&{24.22}&{25.75} & {25.12}\\
         {\textbf{Mismatched AR Prior}} &{\textbf{24.86}}&{\textbf{24.57}}&{\textbf{25.90}} &{\textbf{25.35}}\\
         \hline
        
    \end{tabular*}}
    }
\label{Tab:deconvolution}
\end{table}

\medskip\noindent
Fig.~\ref{Fig:blur} illustrates results on two subsampled images at scale $s=3$. Table \ref{Tab:deconvolution} reports the results over all the test images for 3 different Gaussian blur kernels. Note how regardless of the blur kernel, mismatched AR priors outperform AWGN priors. 

\section{Conclusion}
\label{sec:Con}

\medskip\noindent
This work investigates the robustness of CNN priors trained using the DEQ framework to changes in the measurement operators. To that end, we compare image priors obtained via DEQ to PnP, where image priors are characterized using general AWGN denoisers. We show on two imaging inverse problems that DEQ priors outperform traditional PnP despite DEQ using different measurement operators at training and testing. Our results suggest the robustness of the image priors trained using DEQ to moderate shifts in the measurement operators, thus complementing the recent theoretical analysis of DMBAs under mismatched priors. in~\cite{Shoushtari.etal2022}.


\end{document}